\def \bc {\begin{center}}
\def \ec {\end{center}}

\def \bfr {\begin{flushright}}
\def \efr {\end{flushright}}

\def \v  {\vskip} 

\def \ba {\begin{array}}
\def \ea {\end{array}}

\def \bea {\begin{eqnarray}}
\def \eea {\end{eqnarray}}

\def \be {\begin{equation}}
\def \ee {\end{equation}}
\def \p {\partial}

\def \d {\hbox{d}\,}
\def \square {\hbox{$\sqcup\!\!\!\!\sqcap$}} 
\def \s {\hbox{s}}

\def \e {\hbox{e}}

\def \g {\bar{g}} 
\def \npdos {(\nabla\phi)^2}
\documentstyle[12pt,a4]{article}
\begin{document}
%
\centerline{}
\thispagestyle{empty}
\hfill GR--QC/9706035 
\vskip 3 true cm 

\begin{center} 

{\bf SYMMETRIES IN TWO-DIMENSIONAL DILATON GRAVITY WITH MATTER}  
\vskip5mm 
 Miguel Navarro\footnote{http://www.ugr.es/
$\widetilde{\ }$mnavarro;
 mnavarro@ugr.es} 

\end{center}
\normalsize
\v2mm

{\it Instituto de Matem\'aticas y F\'\i sica Fundamental, 
        CSIC. Serrano 113-123, 28006 Madrid, Spain.}             

\centerline{and}
\vskip 0.5 true cm 
{\it Instituto Carlos I de F\'\i sica Te\'orica y Computacional,
        Facultad  de  Ciencias, Universidad de Granada. 
        Campus de Fuentenueva, 18002, Granada, Spain. }

\vskip10mm

\begin{center}
                        {\bf Abstract}
\end{center}

\footnotesize 

The symmetries of generic 2D dilaton models of gravity 
with (and without) matter are studied in some detail. 
It is shown that $\delta_2$, 
one of the symmetries of the matterless models,  
can be generalized to the case where matter fields of any kind   
are present. The general (classical) solution 
for some of these models, in particular those coupled 
to chiral matter, which generalizes the Vaidya 
solution of Einstein Gravity, is also given. 

\normalsize

\vskip 3mm 



\newpage
\setcounter{page}{1}
\section{Introduction}

 At present, one of the main challenges 
of Theoretical Physics is to devise a quantum theory 
which will provide a complete description of gravity.  
Due to the complexity of the theories involved, 
the Einstein-Hilbert gravity theory for instance, 
this task faces imposing technical 
difficulties. However, equally important or even more serious 
are the conceptual problems that arise. These are not only due  
to a variety of unfamiliar features which are peculiar 
to  diffeomorphism-invariant theories 
but are also due to our present lack of an adequate 
formulation and interpretation of Quantum Mechanics. To an large 
extent, we ignore what a quantum theory {\it really} is -- i.e., we 
ignore what the adjective ``quantum'' really means --  and how we 
should interpret these theories.  Because of that, we ignore not only how to 
produce a quantum theory of gravity but also what would constitute 
a successful completion of this task. 

In this context ``toy theories'' should have 
a crucial concept-clarifying role to play; no wonder  
a variety of them are currently 
under study. Prominent among them are  
the 2D dilaton models of gravity, which are two-dimensional  
general-covariant models whose gravity sector involves, 
along with the space-time metric $g_{\mu\nu}$, a scalar field $\phi$, 
the dilaton (for a review, see Ref. \cite{[Strominger]}). These models, while  
much simpler to handle than are their higher-dimensional cousins, share 
with them not only the conceptual problems which are peculiar 
to diffeomorphism-invariant theories but also their most 
relevant physical features, 
such as the formation of black holes and their subsequent evaporation. 
Some of these models can be obtained, via dimensional reduction,  
from realist, higher-dimensional theories, but this is not a necessary,   
nor even convenient, fact to bear in mind when approaching these 
models, as they may be regarded as something like two-dimensional Brans-Dicke 
theories. 

If attention is restricted to the usual category of models -- that is, 
those with second-order Euler-Lagrange equations of motion --   
a very useful result to take into account is that, excepted isolated 
pathologies which may arise, the actions of these models can all 
be brought, by means of appropiate redefinitions of the 
dilaton field and conformal redefinitions of the metric, 
to the generic form \cite{[Banks],[Gegenberg]} 

\be {S}_{{GDG}}= S_V - S_M\label{GDG}\ee 
where
\be S_V=\int d^2x\sqrt{-{g}}
\left({R}{\phi} + {V} ({\phi})\right) \label{matterlessGDG} \ee 
and $S_M$ is a gravity-matter interaction term which 
may involve the dilaton field as well as the metric . 

Despite their being much simpler that their higher-dimensional 
cousins, few of these models have been solved classically,  
let alone quantum mechanically, when matter fields are present. 
As is well known, solvability, classical or quantum, 
is usually related to the presence of invariances, 
which is the reason that classical 
solvability usually implies quantum solvability. 
Notwithstanding this fact, little attention had been paid to 
the symmetries of these models, apart from the conformal ones 
(see, however, Ref. \cite{[Mann]}). Recently, 
a large variety of symmetries, which are in general 
non-conformal, have been uncovered for the  
matterless 2D dilaton models and they have 
been shown to explain the (classical) 
solvability of these theories \cite{[sym]}. 

In the present paper, we shall consider the generic 
2D dilaton gravity with matter, eq. (\ref{GDG}), 
and shall explore how much of what has been done  
in the matterless case can be extended to theories with matter. 
Nonetheless, to make the present paper as self-contained as possible 
and to make it clear how the symmetries actually 
underly the solvability of these models, 
a brief but systematic view of the matterless theories shall be 
presented in Sect. 2. It is worth discussing beforehand 
some general features of these models. 

The general variation of the Lagrangian in eq. (\ref{GDG}) yields 

\bea 
\delta {\cal L} &=& \sqrt{-g}
\left\{\vphantom{{1\over2}}
\left[ R + V'(\phi)-T_\phi\right]\delta \phi\right. \nonumber \\
&&+\left[g_{\mu\nu}\square\phi -\frac12g_{\mu\nu}V(\phi)
- \nabla_\mu\nabla_\nu\phi-T_{\mu\nu}\right]\delta g^{\mu\nu}\label{deltaGDG}\\
&&+\left({E-L}\right)_A\delta f^A\nonumber\\
&&- \nabla_\alpha s^\alpha\nonumber\eea
where $\nabla_\alpha s^\alpha$ includes all the terms that appear due to 
the ``integrations by part'' which are required to produce the equations 
of motion.  

The equations of motion can be brought to the form: 

\bea 
R+V'(\phi)&=&T_\phi\nonumber\\
\square\phi&=&V+T\label{eqsofmotGDG}\\
\nabla_\mu\nabla_\nu\phi &=& 
\frac12g_{\mu\nu}V+g_{\mu\nu}T-T_{\mu\nu}\nonumber\\
\left({E-L}\right)_A&=&0\nonumber\eea 
where $T$ is the trace of $T_{\mu\nu}$.

Although much of our present study is meant to apply to all 
kinds of matter, we shall exemplify much of our developments 
with a massless scalar field $\xi$, with action 

\be S_M=\frac12\int d^2x\sqrt{-{g}}\,\Omega(\phi)(\nabla\xi)^2\ee
and unspecified function $\Omega$. [The particular case of 
dimensionally reduced spherically symmetric 
Einstein-Hilbert gravity minimally coupled to a 
massless scalar field is recovered with 
$V=2/\sqrt{\phi}$ and 
$\Omega=G\phi$, with $G$ the Newton constant].  

For this example, we have

\bea T_\phi&=&\frac{\Omega'(\phi)}2(\nabla\xi)^2\nonumber\\
T_{\mu\nu}&=&\frac{\Omega}2\left\{\nabla_\mu\xi\nabla_\nu\xi
-\frac12g_{\mu\nu}(\nabla\xi)^2\right\}\label{Txi}\eea

Let us now go back to eq. (\ref{eqsofmotGDG}). It is apparent that 
the second(-from-above) equation is redundant 
as it follows from the third one. However, the same is true for the 
first one. To show this let us first indicate that invariance under 
diffeomorphsims of $S_M$ implies that, when the equations of motion of the 
matter fields are satisfied, the following generalized conservation 
law for the energy-momentum tensor $T_{\mu\nu}$ must be fulfilled  

\be  2\nabla^\nu T_{\mu\nu}+T_\phi\nabla_\mu\phi=0\ee
which implies  

\be T_\phi=-\frac2{\npdos}\nabla^\mu\phi\nabla^\nu T_{\mu\nu}\label{Tphi}\ee 
On the other hand, the affine connection $\nabla_\mu$ and the 
Riemann tensor $R_{\alpha\mu\beta\nu}$ obey 

\be [\nabla_\nu,\>\nabla_\beta]\zeta_\mu =- 
\zeta_\rho {R^\rho}_{\mu\beta\nu} \ee 
for any vector field $\zeta_\mu$. 
Particularizing this equality to $\zeta_\mu=\nabla_\mu\phi$ and
using the fact that in two-dimensions $R_{\mu\nu}=\frac12g_{\mu\nu}R$ yield, 
after a bit of algebra,      

\be R\nabla_\mu\phi +V'\nabla_\mu\phi=-2\nabla^\nu T_{\mu\nu}\ee
which, together with eq. (\ref{Tphi}), yields the desired equality. 

Therefore, all the equations of motion of the theory are 
encapsulated in the last and next-to-last equations 
in eq. (\ref{eqsofmotGDG}). 
As we shall see, this result facilitates  
finding the general classical trajectories. 

Given that in the models we are dealing with 
it is easier to find the conserved currents than the  
associated (Noether) symmetries, 
let us remind the reader, before proceeding,  
of the Noether theorem, suitably taylored for the occasion: 
 Let ${\cal L}={\cal L} \left(\Psi^a\right)$  be an arbitrary 
Lagrangian with general variation $\delta {\cal L}=
\left(E-L\right)_a\delta\Psi^a-\nabla_{\mu}s^{\mu}$. Let  $j_0^\mu$ 
be a current which is conserved on shell and  $\delta_0 \Psi^a$   
a transformation of the fields. 
Then $j_0^\mu$ is the Noether current 
associated to $\delta_0 \Psi^a$ iff, 
without using the equations of motion, 
the following equality holds as an identity:   
\be
\left(E-L\right)_a\delta_0\Psi^a=\nabla_{\mu}j_0^{\mu}\>.
\label{Noethertheorem}
\ee
In general, due to semi-invariance,  
the current $j_0^\mu$ will not equal $s^\mu(\delta_0 \Psi^a)$.

\section{Symmetries and general solution 
for the matterless theories}

With the scalars $\phi$ and $\varphi\equiv\npdos$ and the vector field 
$\nabla_\mu\phi$ scalars, we can built vector and tensor fields 
and we can check whether or not they are conserved. 
Let $J(\phi)$ be a primitive of $V$, $\d J/\d\phi=V$. 
It can be shown that the following results hold: 
 
\noindent{\bf Conserved scalars.} 
The local energy \cite{[Mann],[Gegenberg]}

\be E=\frac12\left(\npdos-J\right), \label{E}\ee 
is a conserved scalar: $\nabla_\mu E=0$, and consequently also is 
$f(E)$ for any function $f$. Moreover, these are 
the only conserved scalars which can be constructed with 
$\phi$ and $\npdos$. 

The Noether symmetry associated to $E$ is given by 
 
\be 
\delta_a \phi=0,\quad
\delta_a g_{\mu\nu}= 
g_{\mu\nu}a_\sigma\nabla^\sigma\phi-\frac12 
\left( a_\mu \nabla_\nu\phi + 
a_\nu \nabla_\mu \phi \right)\label{deltaa}\ee 
with arbitrary constant bivector $a^\mu$. 

\noindent{\bf Conserved currents.} 
The conserved currents of the form 
$j^\mu =A\left(\phi,\>\varphi\right)\nabla^\mu\phi$  
can all be written as follows: 

\be j_{f}^{\mu}=f(E){\nabla^{\mu}\phi\over
\left(\nabla\phi\right)^2}\label{jf}\ee
for some function $f$. The associated symmetries are:

\bea \delta_f\phi&=&0\>,\nonumber\\
\delta_f g_{\mu\nu}&=& -\epsilon f'(E)
\left(g_{\mu\nu}
-{\nabla_{\mu}\phi\nabla_{\nu}\phi\over\left(\nabla\phi\right)^2}
\right) + 
\epsilon f(E)\left({g_{\mu\nu}\over\left(\nabla\phi\right)^2}
-2{\nabla_{\mu}\phi\nabla_{\nu}\phi\over\left(\nabla\phi\right)^4}
\right)\label{deltaf}\eea 
In particular, for $f=1$, the corresponding current and symmetry 
are, respectively,   

\be j_{1}^{\mu}={\nabla^{\mu}\phi\over\left(\nabla\phi\right)^2}\label{j1}\ee

\be \delta_1\phi=0,\quad 
\delta_1 g_{\mu\nu}=
\epsilon\left({g_{\mu\nu}\over\left(\nabla\phi\right)^2}
-2{\nabla_{\mu}\phi\nabla_{\nu}\phi\over\left(\nabla\phi\right)^4}\right)
\label{delta1}\ee

Now, let $j^\mu_R$ be defined by $\nabla_\mu j^\mu_R=R$. 
It is easy to see that the following current is conserved 

\be j_2^{\mu}=j_R^{\mu}+V{\nabla^{\mu}
\phi\over\left(\nabla\phi\right)^2}\label{j2}
\ee
with symmetry

\be \delta_2\phi = \epsilon, \quad 
\delta_2 g_{\mu\nu}=\epsilon  
V\left({g_{\mu\nu}\over\left(\nabla\phi
\right)^2}-2{\nabla_{\mu}\phi\nabla_{\nu}\phi
        \over\left(\nabla\phi\right)^4}\right) 
\label{delta2}\ee

\noindent{\bf Conserved tensors.} 
There exists a great variety of conserved 2-tensors of the form 

\[S^{\mu\nu}= A(\phi,\varphi)\nabla^\mu\phi\nabla^\nu\phi + 
B(\phi,\varphi)g^{\mu\nu}\] 
They are given by the general solution of the equation 
($A_J=\d A/d J,\>A_\varphi=\d A/d\varphi$) 

\[\frac32 A + A_J\varphi +A_\varphi\phi + B_J + B_J=0\]
which implies that $A$ ($B$) can be written as $A=A(J,\varphi)$ 
($B=B(J,\varphi)$). 
For $A=0$, we recover the conserved scalars described above. 
Another useful solution is the following traceless tensor, 
which is basically unique,   

\be S^{\mu\nu}= 
\frac{\nabla^\mu\phi\nabla^\nu\phi}{(\nabla\phi)^4} -
\frac12\frac{g^{\mu\nu}}{\npdos}\label{tracelesstensor}\ee

\subsection{General local solution} 

Once the symmetries are at our disposal, it is easy to solve the equations 
of motion. It can be shown \cite{[sym]} that the following 
``metric'' is invariant under $\delta_2$
 
\[ \g_{\mu\nu}=\frac1{\npdos}\left(g_{\mu\nu}-
\frac{\nabla_\mu\phi\nabla_\nu\phi}{\npdos}\right)\] 
which implies 

\[g_{\mu\nu}=\npdos\g_{\mu\nu} +
\frac{\nabla_\mu\phi\nabla_\nu\phi}{\npdos}\]

As $\nabla^\mu\phi\g_{\mu\nu}=0$, we must have 

\[ \g_{\mu\nu}=Ak_\mu k_\nu\]
with $k^\mu$ the vector density
\be k^\mu=\frac{\epsilon^{\mu\nu}}
{\sqrt{-g}}\nabla_\nu\phi\label{kmu}\ee 

Now, as $\nabla_\mu k_\nu=\nabla_\nu k_\mu$, 
$k_\mu$ is, at least locally, a total derivative: 
$k_\mu=\nabla_\mu t$ for some function $t$. It is then natural to 
choose as local coordinates $r\equiv\phi$ and $t$. 
Now, it is easy to see that $k^\mu$ is a Killing 
vector \cite{[Gegenberg]}, which implies $A=A(r)$.    
The equations of motion (\ref{eqsofmotGDG}) imply $A=-1$. 
Therefore, we finally arrive at the general solution 

\bea \phi&=&r\nonumber\\
 g_{\mu\nu} &=& (2M-J(r))\nabla_\mu t\nabla_\nu t - 
\frac{\nabla_\mu r\nabla_\nu r}{2M-J(r)}\eea 
where $M=-E$ is an arbitrary constant. 

Another elegant way of arriving at the general solutions 
is by using free-field methods. Here we choose the conformal 
gauge 

\be \d s^2=2\e^\rho\d x^+ \d x^-\label{conformalmetric}\ee
for which we have $R= -2\e^{-\rho}\p_+\p_-\rho$ and  
$\square = 2\e^{-\rho}\p_+\p_-$.

In the conformal gauge, the conservation law for the traceless 
tensor in eq. (\ref{tracelesstensor}) takes the form 

\[ 0=\p_+T_{--}=\p_+\left(\frac{\p_-\phi\p_-\phi}{(\nabla \phi)^4}\right),
\qquad 0=\p_-T_{++}=\p_-\left(\frac{\p_+\phi\p_+\phi}{(\nabla \phi)^4}\right)\]
which imply $\frac{\p_+\phi}{(\nabla \phi)^2}=\p_+p,\> 
\frac{\p_-\phi}{(\nabla \phi)^2}=\p_-m$, 
for some $p=p(x^+)$ and $m=m(x^-)$.

A bit of algebra leads us to 

\[\e^\rho=2(\nabla\phi)^2\p_+p\p_-m=2(2E+J)\p_+p\p_-m\]
By fixing the residual gauge as $p=\frac12x^+,
\>m=-\frac12x^-$, we finally arrive at:

\bea\int^\phi {d\tau\over 2M-J\left(\tau\right)}
&=&-\frac12(x^+ -x^-)\nonumber\\
e^\rho&=&\frac12(2M-J)\eea 
 
\section{Symmetries in 2D dilaton gravity with matter}

Let us now introduce matter fields. 
Our most general result in this case is that the  
symmetry (current) $\delta_2$ $(j^\mu_2)$ 
above can be generalized for whatever kind of matter is present. 
More precisely: if $S_M$ is invariant under diffeomorphisms and 
the equations of motion (\ref{eqsofmotGDG}) 
are obeyed, the following current is conserved 

\be  j_2^{\mu}=j_R^{\mu}+V\frac{\nabla^\mu \phi}
{\left(\nabla\phi\right)^2} + 2\frac{\nabla_\nu\phi}
{\left(\nabla\phi\right)^2}T^{\mu\nu}\label{j2matter}
\ee

For us to show this, it suffices to make use of the equations of motion 
(\ref{eqsofmotGDG}), eq. (\ref{Tphi}) and of the fact that 
in two dimension, and for any tensor ${N^\nu}_\mu$, the following 
quantity vanishes identically ($N={N^\rho}_\rho$) 

\[ ({N^\mu}_\alpha -\frac12{\delta^\mu}_\alpha N)({N^\alpha}_\nu
-\frac12{\delta^\alpha}_\nu N)-
\frac12{\delta^\nu}_\mu ({N^\beta}_\alpha 
-\frac12{\delta^\beta}_\alpha N)({N^\alpha}_\beta
-\frac12{\delta^\alpha}_\beta N)\] 
or, equivalently, 

\be {N^\alpha}_\mu {N^\nu}_\alpha=\frac12\delta^\nu_\mu 
 {N^\beta}_\alpha {N^\alpha}_\beta+N{N^\nu}_\mu 
-\frac12\delta^\nu_\mu N^2 \label{algebraicT}\ee

The associated Noether symmetry is 

\bea 
\delta_2\phi &=& \epsilon\\
\delta_2 g_{\mu\nu}&=&\epsilon\left\{ 
V\left({g_{\mu\nu}\over\left(\nabla\phi
\right)^2}-2{\nabla_{\mu}\phi\nabla_{\nu}\phi
        \over\left(\nabla\phi\right)^4}\right) + 
2\frac{T_{\mu\nu}}{(\nabla\phi)^2}-
2T\frac{g_{\mu\nu}}{(\nabla\phi)^2}\right.\label{delta2matter}\\
&&-\left.\frac2{(\nabla\phi)^4}\left(\nabla_\nu\phi 
\nabla^\alpha\phi T_{\alpha\mu} + 
\nabla_\mu\phi \nabla^\alpha\phi T_{\alpha\nu}\right) 
+ 4\frac{g_{\mu\nu}}{(\nabla\phi)^4}\nabla^\alpha\phi\nabla^\beta
\phi T_{\alpha\beta}\right\}\nonumber\eea
The variation $\delta f^A$ of the matter fields 
is such that off-shell 

\be \left({E-L}\right)_A\delta f^A= 2\frac{\nabla_\nu\phi}
{\left(\nabla\phi\right)^2}\nabla_\mu T^{\mu\nu}\ee  
For our exemplifying scalar matter field, 
this transformation is 

\be \delta\xi = \epsilon\frac{\nabla^\mu\phi}{(\nabla\phi)^2}\nabla_\mu\xi = 
\epsilon\frac{\nabla\phi\nabla \xi}{(\nabla\phi)^2}\ee
It is easy to see that, in spite of what happens in the matterless 
case, this symmetry does not correspond to the diffeomorphism 
generated on-shell by the vector field $s^\mu = 
\frac{\nabla^\mu\phi}{(\nabla\phi)^2}$. 

Let us now consider currents of the form

\be S^\mu=A(\phi,\>\varphi)\nabla^\mu\phi+ 
B(\phi,\>\varphi)\nabla_\nu T^{\mu\nu}\ee 
Conservation implies 

\bea 0&=&A_\phi \npdos +A_\varphi V\npdos +AV 
-\frac12B\npdos T_\phi\nonumber\\
&&+\left[ 2A_\varphi\npdos  +A +\frac12BV \right] T \nonumber\\  
&&+\left[ B+B_\varphi\npdos\right] \left(T^2-
T_{\alpha\beta}T^{\alpha\beta}\right)\\
&&+\left[-2A_\varphi +B_\phi 
+B_\varphi V\right]\nabla_\mu\phi\nabla_\nu\phi T^{\mu\nu}
\nonumber\eea
where eq. (\ref{algebraicT}) has been used. 
It follows that 

\noindent 1) If $T_\phi=0,\>T=0$ and $V'=\beta V$ 
(with $\beta=$ constant)  the following current is conserved 

\be S^\mu_\beta =-\beta \nabla^\mu\phi + 
V \frac{\nabla^\mu \phi}{\left(\nabla\phi\right)^2} + 
2\frac{\nabla_\nu \phi}{\left(\nabla\phi\right)^2}T^{\mu\nu}\ee 
The Noether symmetry associated to $j^\mu_\beta=j^\mu_2-S^\mu_\beta$ 
is conformal and is given by \cite{[sym]}

\be \delta_\beta\phi=\epsilon,\quad 
\delta g_{\mu\nu}=-\epsilon\beta g_{\mu\nu},\quad \delta f^A=0
\label{deltabeta}\ee

\noindent 2) If $T_\phi=0,\>T=0$ and 
$T_{\alpha\beta}T^{\alpha\beta}=0$ (chiral matter) 

\be S^\mu_f =f(E)\nabla_\nu\phi T^{\mu\nu}\ee 
is a conserved current for any function $f$. 
The associated symmetry transformation of the gravity sector is 

\be\delta\phi=0,\qquad \delta g_{\mu\nu}= T_{\mu\nu} \ee 

\noindent{\bf Conserved 2-tensors.}
Consider now (symmetric) tensor fields of the form 

\be S^{\mu\nu}= A(\phi,\varphi)\nabla^\mu\phi\nabla^\nu\phi +
B(\phi,\varphi)g^{\mu\nu} + C(\phi,\varphi)T^{\mu\nu}\ee 
It can be show that 

\noindent 1) For $T_\phi=0$ and $T=0$ the tensor fields with 

\bea C&=&C(\phi),\quad \hbox{$C$ an arbitrary function}\nonumber\\
B&=&(C_\phi +\alpha)\varphi +2\alpha J +\frac12\int^\phi VC_\phi\\
A&=&-C_\phi -2\alpha\nonumber\eea
are conserved. If $T\neq0$, these tensors are still conserved if   
$\alpha=0$. 
The tensor with $C=\phi,\>\alpha=-\frac12$, 

\be  J^{\mu\nu}=\frac12g^{\mu\nu}\left((\nabla\phi)^2 
-J(\phi)\right) +\phi T^{\mu\nu}\label{energymatter}\ee
can be regarded as the generalization of the local energy 
of the massless models. 
The conserved tensor with $C=0$ is purely kinematical -- it 
does not depend on the matter fields: 

\be S_0^{\mu\nu}=\nabla^\mu\phi\nabla^\nu\phi -
\frac12g^{\mu\nu}\left(\nabla\phi\right)^2 -g^{\mu\nu}J\ee

\subsection{Chiral matter}

We define chiral matter as the one whose energy-momentum tensor obeys 

\be T_\phi=0,\qquad {T^\mu}_\mu=0 
\quad \hbox{and}\quad T_{\mu\nu}T^{\mu\nu}=0\ee 
In the conformal gauge we have 

\be 0=T_{\mu\nu}T^{\mu\nu}=2\e^{-2\rho}T_{++}T_{--}  \ee 
Therefore the set of all solution splits in two sectors of 
left-moving  and right-moving fields. Let us choose $T_{--}=0$. 
The conservation law for the tensor in 
eq. (\ref{energymatter}) implies: 

\bea \p_-E &=&0\quad\Longrightarrow E=P(x^+)\nonumber\\
\p_+ E + \nabla^+\phi T_{++}&=&0\quad\Longrightarrow 
\nabla^+\phi=p(x^+)\eea
Thus, we are led to the equations

\bea e^{-\rho}\p_-\phi&=&p(x^+)\nonumber\\
\p_+\phi&=&\frac{P}{p} +\frac1{2p}J\\
\frac{\p_+P}{p}&=&-T_{++}\nonumber\eea

However, it does not appear that, in the present coordinates,
 these equations can be solved in general for arbitrary 
$J$ and $T_{++}$. To go further, 
let us make a change of coordinates 

\be x^+=u, \qquad x^-=x^-(u,r)\ee 
so that $\phi\equiv r$. The metric and the energy-momentum tensor 
in this coordinates takes the form

\bea \d\s^2&=&e^\lambda\left(2\d\phi\d u+e^\lambda A(\d u)^2\right)\nonumber\\
T&=&T_{uu}(du)^2\eea 
with $T_{uu}=T_{uu}(u)$. 
Now, by writting down in this gauge the equations of motion 
(\ref{eqsofmotGDG}), it is easy to show that $\e^\lambda$ 
is pure gauge, and finally arrive at the following general solution 

\be \d\s^2= 2\d r \d u + (2M(u)-J) (\d u)^2\ee 
with  

\be M(u)=\int^u \d\tilde{u} T_{uu}(\tilde{u})\label{Mu}\ee
This metric generalizes, for arbitrary  potential $V$,  
the Vaidya solution of Einstein gravity.  

\section{Conclusions}

We have analysed in a rather systematic way the conserved currents 
and the symmetries of the 2D dilaton models 
of gravity with and without matter. 
In particular we have shown that $\delta_2$ 
can be extended to models coupled to any kind of matter. 
We have also shown how analytic solvability is directly related with 
the existence of invariances -- and hence of conserved currents. 
In fact, almost all the models 
which are known to be solvable fall into one of the categories of 
symmetric models that we have described in the present paper. 
The Jackiw-Teitelboim model, $V=4\lambda^2\phi$, coupled 
to conformal matter may be regarded as an exception 
to this rule,  as it is solvable \cite{[canonicaltr]} even though, 
apart from $\delta_2$, it is not known to have any 
additional symmetry. This model is solvable because, 
in the conformal gauge, the first equation 
in eq. (\ref{eqsofmotGDG}) is a Liouville equation.  
This Liouville equation, despite being a limiting case of the free-field 
equation of the exponential model which is associated to the symmetry in eq. 
(\ref{deltabeta}) (see also Ref. \cite{[sym]}),   
does not appear to be related itself to any invariance.
In fact, the whole problem of solving the generic 2D dilaton 
models of gravity (with conformal matter) 
can be regarded as a generalization of Liouville theory. 
It would be interesting to see 
if the machinery of integrable systems is useful here. 

The massless 2D dilaton models have been shown 
to be related to Poisson-$\sigma$-models, and this  
seems to explains their highly symmetric nature 
(see, for instance, Ref. \cite{[Kummer],[Poisson]} 
and references therein).  On the other hand, 
the underlying reason why $\delta_2$ is so general 
remains a mistery to us as of the writing of this paper. It seems 
to indicate that these models have a degree of unity   
previously unexpected. This unity may have important 
consequences in relation to their solvability. 

\section*{Acknowledgements}
The author has profitted from discussions with 
J. Cruz and J. Navarro-Salas. 
He acknowledges the Spanish MEC, CSIC and IMAFF (Madrid)
for a research contract. 

This work was partially supported by the 
Comisi\'on Interministerial de Ciencia y Tecnolog\'{\i}a 
and DGICYT.

\end{document}